# Compressive spectral image classification using 3D coded convolutional neural network

Hao Zhang,[1] Xu Ma,[1,] * Xianhong Zhao,[1] Gonzalo R. Arce[2]

[1]*Key Laboratory of Photoelectronic Imaging Technology and System of Ministry of Education of China, School of Optics and Photonics, Beijing Institute of Technology, Beijing 100081, China*
[2]*Department of Electrical and Computer Engineering, University of Delaware, Newark, Delaware 19716, USA*
*\*Corresponding author: maxu@bit.edu.cn*

**Abstract:** Hyperspectral image classification (HIC) is an active research topic in remote sensing. Hyperspectral images typically generate large data cubes posing big challenges in data acquisition, storage, transmission and processing. To overcome these limitations, this paper develops a novel deep learning HIC approach based on compressive measurements of coded-aperture snapshot spectral imagers (CASSI), without reconstructing the complete hyperspectral data cube. A new kind of deep learning strategy, namely 3D coded convolutional neural network (3D-CCNN) is proposed to efficiently solve for the classification problem, where the hardware-based coded aperture is regarded as a pixel-wise connected network layer. An end-to-end training method is developed to jointly optimize the network parameters and the coded apertures with periodic structures. The accuracy of classification is effectively improved by exploiting the synergy between the deep learning network and coded apertures. The superiority of the proposed method is assessed over the state-of-the-art HIC methods on several hyperspectral datasets.

## 1. Introduction

Hyperspectral imaging acquires hundreds image planes spanning wavelengths from the visible to the infrared wavelengths. The rich spectral information of hyperspectral images has been widely employed in a range of remote sensing applications, such as ecological science, geological science, hydrological science, and precision agriculture [1, 2]. Hyperspectral image classification (HIC) technology plays a crucial role in these applications, where a label is assigned to each spatial pixel of the scene based on its spectral signature. A large number of HIC methods have been proposed based on k-nearest-neighbors, maximum likelihood criterion, logistic regression, and support vector machine (SVM) [3-6]. Over the past several years, deep learning has become one of the most efficient signal processing approaches with great potential in hyperspectral imaging and classification [7-9].

Traditional HIC methods need three-dimensional (3D) spatio-spectral datasets, which are captured by whisk broom or push broom scanning spectral imaging systems [10, 11]. However, these systems require a time-consuming scanning process in the spatial or spectral domain, and lead to a big data size to be stored, transmitted, and processed. To overcome these limitations, Wagadarikar et al. introduced the concept of coded aperture snapshot spectral imaging (CASSI) system based on the compressive sensing (CS) theory [12]. The CASSI system simultaneously senses and compresses the 3D spectral data cube with just a single or a few two-dimensional (2D) projection measurements [13-16]. The complete 3D spectral data cube can be reconstructed from the compressive measurements, which can then be used for classification and processing. However, spectral image classification based on CASSI is a challenging task since the reconstruction procedure is very time-consuming and noise sensitive.

Recently, a supervised compressive spectral image classifier was proposed for CASSI system, where the hyperspectral pixel was approximately represented as a sparse linear combination of samples in an overcomplete training dictionary [17]. The sparse coefficients were recovered from a set of CASSI compressive measurements to determine the clusters of the unknown pixels. Although this method does not need to reconstruct the complete 3D spectral data cube, recovering the sparse coefficients for all hyperspectral pixels is still computationally intensive. In addition, to improve the accuracy of compressive spectral image classifiers, two improvements in the compressive classifier system were made. One is the measurement stage and the other is classification stage. In the measurement stage, the coded apertures were optimized based on the restricted isometry property (RIP), which is a widely used criterion to obtain the optimal reconstruction performance in CS theory [17]. In the classification stage, the sparse dictionary was optimized and different sparsity-based classifiers were proposed [18-20]. However, the coded apertures with optimal reconstruction performance are not necessarily the best to achieve the highest classification accuracy, since the reconstruction itself may introduce unexpected artifacts that are not supported by the compressive measurements. In addition, current methods ignored the codependency between the measurement stage and classification stage, which limits the improvement of classification accuracy. Recently, the combination of optics and deep learning has become a trend, and some end-to-end optimization approaches of optics and image processing were proposed [21-23]. This paper proposed a novel deep learning approach, namely 3D coded convolutional neural network (3D-CCNN) that efficiently solves the HIC problem directly from the compressive domain without reconstruction, and jointly optimizes the coded apertures.

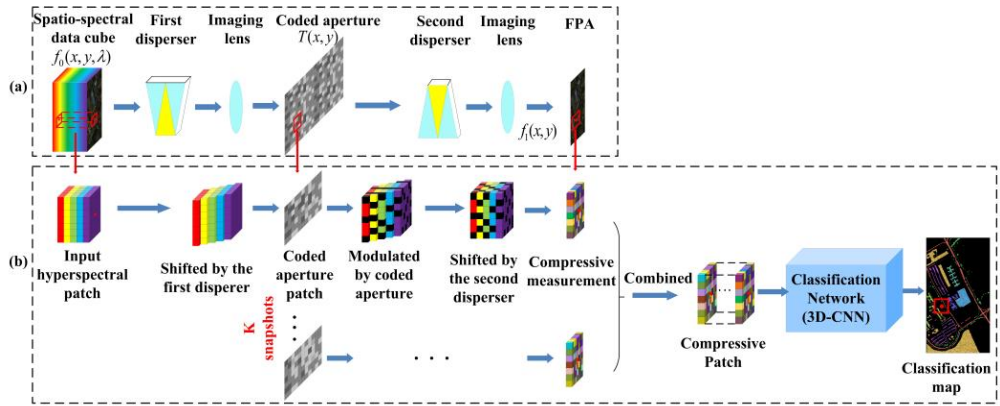

Fig. 1. Sketches of (a) the DD-CASSI system and (b) the proposed 3D-CCNN framework. The DD-CASSI is used in the measurement stage, and the 3D-CNN is used in the classification stage. The imaging model of DD-CASSI can be decomposed into patch-based models to keep the dimensionality consistent with the classification network. The 3D-CCNN system combines the coded aperture optimization and the HSI classification into a single framework.

As shown in Fig. 1(a), the CASSI system with a dual-disperser architecture (DD-CASSI) is used in the measurement stage to capture the compressive measurement of a target scene. In DD-CASSI, the hyperspectral data cube is first shifted by the front dispersive element, then modulated by a coded aperture in spatial domain, and finally shifted back using the second dispersive element [14]. After that, the encoded hyperspectral data cube is projected onto a 2D integrated detector. In the measurement process, we switch different coded apertures to capture a set of compressive measurements. Different from the single-disperser-based CASSI system [12, 13], the compressive measurements in DD-CASSI have the same spatial dimensions as the target scene. Each detector element receives the information from all spectral bands with different codes. These characteristics enable us to decompose the imaging model of DD-CASSI into patch-based models, whose dimensionality is consistent with the following classification network, since the classification is implemented in a patch-based manner.

As shown in Fig. 1(b), the classification stage consists of a 3D convolutional neural network (3D-CNN) to predict the classification map directly in the compressive domain, without reconstructing the complete 3D data cube. Given the correlation of hyperspectral data across both spatial and spectral dimensions, the 3D-CNN takes the compressive measurement patches as the input. In order to obtain the optimal coding from a small training subset of the hyperspectral data, the coded apertures are designed as periodic patterns to reduce the independent optimization variables. Thus, we only need to optimize one period of coded apertures, and then periodically extend it to the entire coded pattern. Taking the full advantages of the patch-based model, an end-to-end training method is proposed to jointly optimize the coded apertures and the classification network parameters. In this work, the coded aperture optimization and hyperspectral image classification are concatenated into a single system, dubbed 3D-CCNN, which effectively increases the degrees of optimization freedom and improves the classification accuracy.

The main contributions of this paper are twofold. First, we integrate deep learning with CASSI to solve for the classification problem directly in the compressive domain, thus avoid the time-consuming reconstruction procedure and alleviate the influence of reconstruction artifacts. Second, the hardware-based coded aperture and software-based classification network are unified into one framework, coined 3D-CCNN. This paper is thus bridges the gap between coded aperture design and classification to increase the degrees of optimization freedom. Then, the end-to-end training method is used to jointly optimize the coded apertures and network parameters, which effectively improves the classification accuracy. The superiority of the proposed method over some of the state-of-the-art approaches is verified by a set of simulations. This is the updated version of the articles arXiv:2009.11948v1 [eess.IV] and arXiv:2009.11948v2 [eess.IV], which are respectively entitled "Compressive Spectral Image Classification using 3D Coded Neural Network" and "Joint coded aperture optimization and compressive hyperspectral image classification using 3D coded neural network". This updated version includes several improvements over the previous versions. The simulation results have been corrected and updated, and the method to generate the coded apertures has been modified. Several data sets, images, and descriptions have also been improved and updated. The contributions in the paper and the modifications in the revision have been provided by the corresponding author Xu Ma, and all of the coauthors.

## 2. Forward imaging model of DD-CASSI system

As shown in Fig. 1(a), DD-CASSI system employs two opposite dispersers and a coded aperture to encode the hyperspectral data cube in both, the spatial and spectral domains [14]. Let $f_0(x, y, \lambda)$ be the hyperspectral data cube of the target scene, where $x$ and $y$ are the spatial coordinates, and $\lambda$ is the spectral coordinate. The hyperspectral data cube is first laterally shifted as a function of wavelength by the front disperser to form the skewed data cube, which is then projected by an imaging lens onto the coded aperture plane. The skewed data cube is modulated in the spatial domain by the coded aperture whose transmission function is denoted by $T(x, y)$. Subsequently, the coded source planes are shifted back into a standard cube by the second disperser, and integrated along the $\lambda$ axis on the 2D focal plane array (FPA) detector. The dispersion effect enables the coded aperture to introduce distinguishable spatial modulations in different spectral bands. The measurement intensity on FPA detector can be formulated as [14]:

$$Y(x, y) = \int T(x - \alpha(\lambda - \lambda_c), y) f_0(x, y, \lambda) d\lambda , \qquad (1)$$

where $\alpha$ and $\lambda_c$ are the linear dispersion rate and the center wavelength of prisms, respectively.

Due to the pixelated nature of detector array, the continuous model in Eq. (1) can be transformed into a discrete form. Suppose we take $K$ snapshots in total with different coded apertures, and $\mathbf{T}^k$ represents the coded aperture pattern used in the $k$th snapshot. Then, the $k$th snapshot measurement on FPA is given by

$$\mathbf{Y}_{ij}^k = \sum_{l=0}^{L-1} \mathbf{F}_{i,j,l} \mathbf{T}_{i,j+l}^k + \boldsymbol{\omega}_{i,j}^k, \qquad (2)$$

where $i$ and $j$ are the pixel coordinates in spatial domain, and $l$ is the pixel coordinate in spectral domain; $\mathbf{F}$ is the 3D hyperspectral data cube of target with dimension $N \times M \times L$; $\mathbf{F}_{i,j,l}$ is the voxel at the spatial coordinate $(i,j)$ in the $l$th spectral band; and $\boldsymbol{\omega}_{i,j}^k$ is the measurement noise on the detector; $\mathbf{Y}^k$ is the compressive measurement with dimension $N \times M$; and the dimension of $\mathbf{T}^k$ is $N \times (M+L-1)$. Next, we transform Eq. (2) into a matrix form. Let $\mathbf{y}^k \in R^{NM \times 1}$ and $\mathbf{f} \in R^{NML \times 1}$ be the vectorized representations of $\mathbf{Y}^k$ and $\mathbf{F}$, respectively. Then, we have

$$\mathbf{y}^k = \mathbf{H}^k \mathbf{f} + \boldsymbol{\omega}^k, \qquad (3)$$

where $\mathbf{H}^k$ is the system matrix representing the effect of the $k$th coded aperture and the dispersers, and $\boldsymbol{\omega}^k$ is the vector of measurement noise. Taking into account all of the $K$ snapshots, the measurements can be concatenated together, and the forward imaging model becomes [15, 24, 25]:

$$\mathbf{y} = \mathbf{H}\mathbf{f} + \boldsymbol{\omega}, \qquad (4)$$

where $\mathbf{y} = [(\mathbf{y}^1)^T, (\mathbf{y}^2)^T, ..., (\mathbf{y}^K)^T]^T$ and $\mathbf{H} = [(\mathbf{H}^1)^T, (\mathbf{H}^2)^T, ..., (\mathbf{H}^K)^T]^T$. Suppose the data cube is highly correlated across the spatial and spectral domains, and is sparse in some representation basis $\boldsymbol{\Psi}$ [26-28]. Then, $\mathbf{f}$ in Eq. (4) can be represented as $\mathbf{f} = \boldsymbol{\Psi}\boldsymbol{\theta}$, where $\boldsymbol{\Psi} = \boldsymbol{\Psi}_1 \otimes \boldsymbol{\Psi}_2$ is a 3D representation basis, $\boldsymbol{\Psi}_1$ is the 2D wavelet Symmlet-8 basis to depict the correlation in spatial domain, $\boldsymbol{\Psi}_2$ is the one-dimensional (1D) DCT basis in spectral domain, $\otimes$ is the Kronecker product, and $\boldsymbol{\theta}$ is the coefficient vector. Substituting $\mathbf{f} = \boldsymbol{\Psi}\boldsymbol{\theta}$ into Eq. (4), we have

$$\mathbf{y} = \mathbf{H}\boldsymbol{\Psi}\boldsymbol{\theta} + \boldsymbol{\omega}. \qquad (5)$$

It is noted that the matrix $\mathbf{H}$ is sparse and highly structured, which includes a set of diagonal line structures determined by the coded aperture entries $\mathbf{T}_{i,j+l}^k$. An illustrative example of the matrix $\mathbf{H}$ is shown in Fig. 2, where $K=2$, $N=M=6$, $L=3$, and the coded aperture patterns obey the Bernoulli distribution with 50% transmittance.

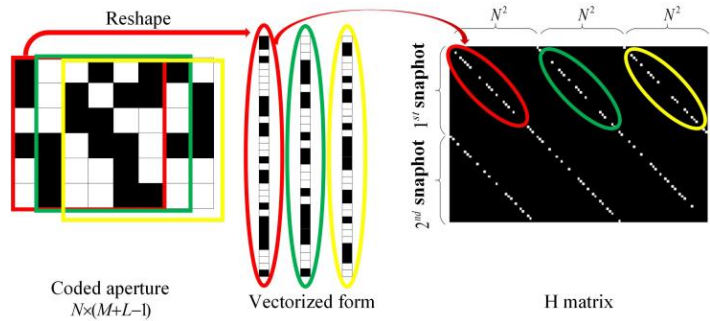

Fig. 2. An illustrative example of the matrix $\mathbf{H}$ for the Bernoulli random coded apertures, where $K=2$, $N=M=6$, $L=3$.

## 3. 3D-CCNN approach for hyperspectral image classification

In this section, we build up a seven-layer 3D-CNN to solve the classification problem directly in the compressive domain. Then, we decompose the forward imaging model of the DD-CASSI into patch-based models, and the periodic design of coded apertures is introduced. The coded aperture and classification network are further connected into a uniform framework, namely 3D-CCNN. The joint training method of 3D-CCNN is presented at the end of this section. The sketch of the 3D-CCNN framework is shown in Fig. 3.

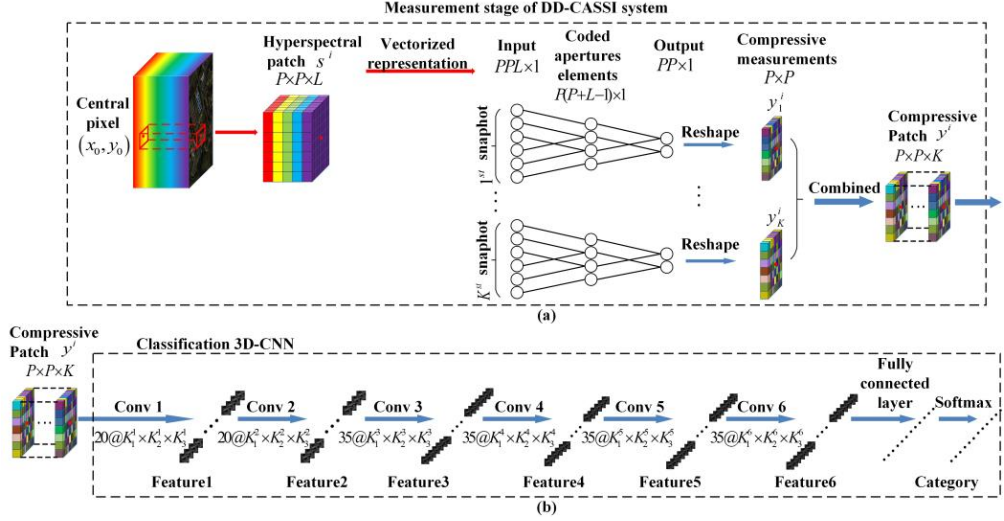

Fig.3 Sketch of the 3D-CCNN framework, which connects (a) the measurement stage of DD-CASSI system and (b) the 3D-CNN classification network. The coded apertures and classification network are jointly trained in an end-to-end supervised manner.

*3.1 Compressive spectral image classification using 3D-CNN*

The DD-CASSI system is used to acquire several compressive measurements with different coded apertures. Initially, random coded apertures were used in DD-CASSI. Assume the hyperspectral data cube of target scene consists of $N \times M$ spatial pixels and $L$ spectral bands. Taking $K$ snapshots, we can obtain a compressive measurement data cube with dimension of $N \times M \times K$. The goal is to solve the HIC problem directly from the compressive measurements without reconstruction.

Recently, deep learning has been proved to render accurate semantic interpretation of the underlying datasets [29]. Given the 3D nature of the compressive measurement data cube, the 3D-CNN framework is chosen to perform the classification task, since it can simultaneously exploit information from all measurement slices with different coding, which is essential to improve classification performance [8, 9, 29-31].

As shown in Fig. 3, the HIC problem is pixel-based, where each spatial pixel on the hyperspectral images is associated with a specific classification label. Note that the pixels inside a small neighborhood often reflect relevant information of the underlying objects or materials. Thus, the information of measurement data surrounding a pixel is helpful to improve the classification accuracy of that pixel. For each pixel under consideration, we truncate a small patch around it from the compressive measurement data cube. The dimension of the patch is $P \times P \times K$, where $P \times P$ is the spatial size, and $K$ is equal to the number of compressive measurements. The dimension $P$ is often chosen as an odd number to keep the symmetry. The center of the patch is located on the pixel to be classified. Then, the patch is used as the input of the 3D-CNN, and the output is the classification label of the central pixel. Next, we describe the structure of the 3D-CNN in more detail.

Talking about the depth and width of the 3D-CNN is a very rich debate that generates a lot of questions. However, it has been recently proved that one of the keys for better performances is to find the right balance between the network's depth and width [29]. To harmonize the cost and accuracy of a deep network, the 3D-CNN built up in this paper consists of 7 layers, including 6 convolutional layers and 1 fully connected layer. In addition, the first and second layers are characterized with 20 filters, whereas the rest of the layers have 35 filters. As shown in Fig.3 (b), the convolutional layers transfer the input data to a series of 3D feature maps, which are gradually reduced into a 1D feature vector. The 1D feature vector is inputted to a fully-connected layer, the output of which is then fed into a Softmax classifier to calculate the classification result. From the first layer to the sixth layer, the dimensions of the 3D convolution kernels are $20\times3\times3\times3$ (i.e., $K_1^1=3$, $K_2^1=3$, and $K_3^1=3$), $20\times3\times1\times1$ (i.e., $K_1^2=3$, $K_2^2=1$, $K_3^2=1$), $35\times3\times3\times3$ (i.e., $K_1^3=3$, $K_2^3=3$, and $K_3^3=3$), $35\times3\times1\times1$ (i.e., $K_1^4=3$, $K_2^4=1$, and $K_3^4=1$), $35\times3\times1\times1$ (i.e., $K_1^5=3$, $K_2^5=1$, and $K_3^5=1$), and $35\times2\times1\times1$ (i.e., $K_1^6=2$, $K_2^6=1$, and $K_3^6=1$), respectively. For instance, the "$20\times3\times3\times3$" means that there are twenty 3D-kernels with dimension of $3\times3\times3$ (i.e., two spatial dimensions and one spectral dimension). Denote $W$ as the parameter set of the 3D-CNN, including all the convolution kernels, weights and biases.

*3.2 Patch-based model with periodic coded apertures*

To keep the consistence of the dimensionality, we first decompose the forward imaging model of DD-CASSI into patch-based models. As shown in Fig. 4, we first divide the hyperspectral data cube and compressive measurements in Eq. (4) into small patches. Let $y^i$ with dimension $P\times P\times K$ be the *i*th measurement patch truncated from the compressive measurement data cube. Define $y_k^i \in R^{P\times P}$ as the *k*th slice of $y^i$. Then, we can trace $y_k^i$ from the detector back through the DD-CASSI system, and find out the corresponding 3D patch $s^i$ in the original hyperspectral data cube $\mathbf{F}$. The patch $s^i$ is a $P\times P\times L$ data cube, where $L$ denotes the number of spectral bands. Due to the first disperser, the hyperspectral patch $s^i$ is shifted into a parallelepiped, and then modulated by a coded aperture patch $t_k^i$ with dimension of $P\times(P+L-1)$, where $t_k^i$ represents the coded aperture patch associated with $s^i$ at the *k*th snapshot.

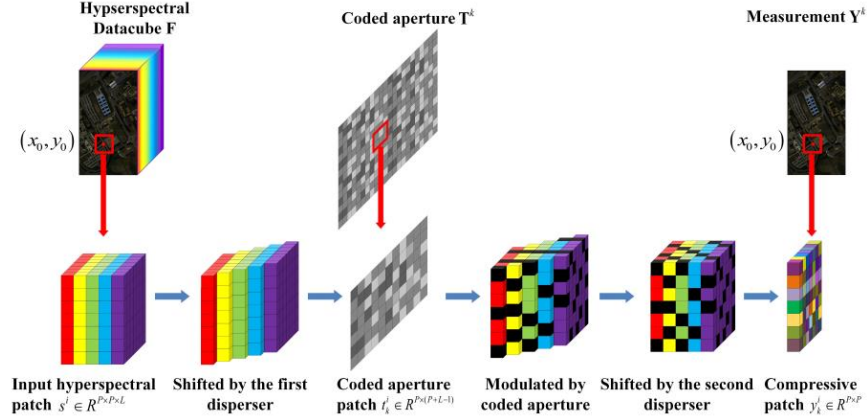

Fig. 4. The patch-based imaging model of DD-CASSI. For each snapshot, a 3D hyperspectral patch $s^i \in R^{P \times P \times L}$ corresponds to a compressive measurement patch $y_k^i \in R^{P \times P}$ on the detector. The hyperspectral patch $s^i$ is modulated by a coded aperture patch $t_k^i \in R^{P \times (P+L-1)}$.

To describe in detail, each spectral band of $s^i$ is modulated by different coding templates with dimension $P \times P$ due to the dispersive effect, and every coding template can be regarded as a part of the coded aperture patch $t_k^i$. Denote the central point of $y_k^i$ as $Y_{x_0,y_0}^k$, where $(x_0, y_0)$ is the central coordinate. Then, the central pixel of the hyperspectral patch $s^i$ is $F_{x_0,y_0,l}$, where $l$ is the spectral coordinate. The $l$th spectral band of $s^i$ and the coded aperture patch $t_k^i$ can be given by:

$$s_l^i = \begin{bmatrix} F_{x_0-q,y_0-q,l} & \cdots & F_{x_0-q,y_0+q,l} \\ \vdots & F_{x_0,y_0,l} & \vdots \\ F_{x_0+q,y_0-q,l} & \cdots & F_{x_0+q,y_0+q,l} \end{bmatrix}, \text{ and } t_k^i = \begin{bmatrix} T_{x_0-q,y_0-q}^k & \cdots & T_{x_0-q,y_0+q+L-1}^k \\ \vdots & \ddots & \vdots \\ T_{x_0+q,y_0-q}^k & \cdots & T_{x_0+q,y_0+q+L-1}^k \end{bmatrix}, \quad (6)$$

where $q = \lfloor P/2 \rfloor$, and $\lfloor \ \rfloor$ is the rounding operator.

Let $F_{m,n,l}$ be the $(m,n)$th pixel in the $l$th spectral band of $s^i$, and let $Y_{m,n}^k$ be the $(m,n)$th pixel in the $k$th band of $y^i$. The spatial dimension of input patch $s^i$ and the measurement patch $y^i$ is the same. According to Eq. (6), the relationships between the subscripts are $m = x_0 + \alpha$, $n = y_0 + \alpha$, and $\alpha \in [-q, q]$. Then, we have

$$Y_{m,n}^k = \sum_{l=0}^{L-1} F_{m,n,l} T_{m,n+l}^k, \quad (7)$$

where $T_{m,n+l}^k (l=0,...,L-1)$ is the $(m, n+l)$th pixel of the coded aperture patch $t_k^i$.

When taking $K$ snapshots, there are totally $KN(M+L-1)$ coded aperture entries to be optimized. If all the coded aperture variables are independent on each other, it is impossible to train them by only using a small set of training samples on the hyperspectral data cube. That is because the training problem will become underdetermined if the number of variables exceeds the number of training samples. To circumvent this problem, we design the coded apertures with periodic patterns, where each coded aperture is cyclically filled by a basic block with dimension $B \times B$, and different coded apertures have different basic blocks. Denote $C^k \in R^{B \times B}$ as the basic block for the $k$th coded aperture $T^k \in R^{N \times (M+L-1)}$. The pixel value of $T_{m,n+l}^k$ is equal to $\mathbf{C}^{k}_{m_1,n_1}$, where $m_1 = m\%B$, $n_1 = (n+l)\%B$, and $\%$ is the remainder operation. Then, Eq. (7) can be rewritten as

$$\mathbf{Y}^{k}_{m,n} = \sum_{l=0}^{L-1} \mathbf{F}_{m,n,l} \mathbf{C}^{k}_{m_1,n_1} . \tag{8}$$

where $m_1 = (x_0+k)\%B$, $n_1 = (y_0+k+l)\%B$, and $k \in [-q,q]$.

Taking all of the $K$ snapshots into account, the entries of $\mathbf{C}^{k}_{m_1,n_1}$ can be concatenated together to form:

$$\mathbf{C}(i) = \left[ \mathbf{C}^{1}_{m_1,n_1}, \mathbf{C}^{2}_{m_1,n_1}, ..., \mathbf{C}^{k}_{m_1,n_1}, ..., \mathbf{C}^{K}_{m_1,n_1} \right], \text{ for different } m_1 \text{ and } n_1, \tag{9}$$

where $i$ indicates the input patch $s^i$. The set $\mathbf{C}(i)$ consists of all the entries in the $K$ basic blocks associated with $s^i$. As shown in Fig. 3(a), the coded aperture can be regarded as a pixel-wise connected layer to encode the input data cube and obtain the measurement patches, which are then inputted to the 3D-CNN proposed in Section 3.1. The effect of coded apertures is equivalent to a virtual connected layer in the 3D-CCNN framework. Then, we can jointly optimize the basic blocks of coded apertures and other network parameters using an end-to-end supervised training method.

### 3.3 Joint training method of 3D-CCNN

This section proposes an end-to-end supervised training method for 3D-CCNN. The set of all parameters to be optimized is denoted as $\Theta = [W, \mathbf{C}(i)]$, where $W$ represents the parameters in the seven-layer 3D-CNN model described in Section 3.1, and $\mathbf{C}(i)$ represents the parameters of coded apertures defined in Eq. (9). In the hyperspectral data cube, we randomly choose 30% labeled samples as the training data, and the remaining 70% pixels are used for testing.

In this work, the softmax loss is used as the objective function in the training process. Suppose the input of softmax classifier is a vector $x \in R^{M \times 1}$, where $M$ is the number of classes. The output of softmax classifier is an $M \times 1$ vector $p = [p_1, p_2, ..., p_M]$. Then, the loss function is

$$\text{Loss} = -\sum_{j=1}^{M} l_j \log(p_j) , \tag{10}$$

where $l = [l_1, l_2, ..., l_M]$ is an $M \times 1$ true label vector. For $i \in [1, M]$, the $l_j = 1$ if the pixel under consideration actually belongs to the $j$th class, otherwise $l_j = 0$. The $p_j$ is the $j$th element of the output vector, which represents the probability that the pixel belongs to the $j$th class:

$$p_j = e^{x_j} / \sum_{i=1}^{M} e^{x_i} , \tag{11}$$

where $x_i$ and $x_j$ represent the $i$th and the $j$th elements of the input vector $x$, respectively. The loss function is minimized using back propagation method. The network parameters are updated as:

$$\{W^{v+1}, \mathbf{C}(i)^{v+1}\} = \{W^v, \mathbf{C}(i)^v\} - \eta \cdot \nabla Loss\{W^v, \mathbf{C}(i)^v\} , \tag{12}$$

where $v$ indicates the iteration number, $\eta$ is the learning rate, and $\nabla Loss\{W^v, \mathbf{C}(i)^v\}$ represents the gradient of the loss function with respect to the variables. After the training process, the leaned basic block $\mathbf{C}^k$ can be tiled to form the complete coded aperture $\mathbf{T}^k$.

In this paper, the coded apertures are greyscale. At the end of the training process, we use a function to limit the value of coded apertures between 0 and 1, i.e.

$$\mathbf{C}^k = \begin{cases} 1, & \mathbf{C}^k > 1, \\ 0, & \mathbf{C}^k < 0 \end{cases}. \tag{13}$$

As shown in Fig. 5, the blue arrows represent the end-to-end training process, and the red arrows represent the testing process. In the testing process, the optimized coded apertures obtained by training method are first manufactured and installed in the DD-CASSI system. A set of compressive measurements are captured by the detector. Subsequently, the compressive measurement data cube is decomposed into patches, which are then inputted into the 3D-CNN to obtain the classification results.

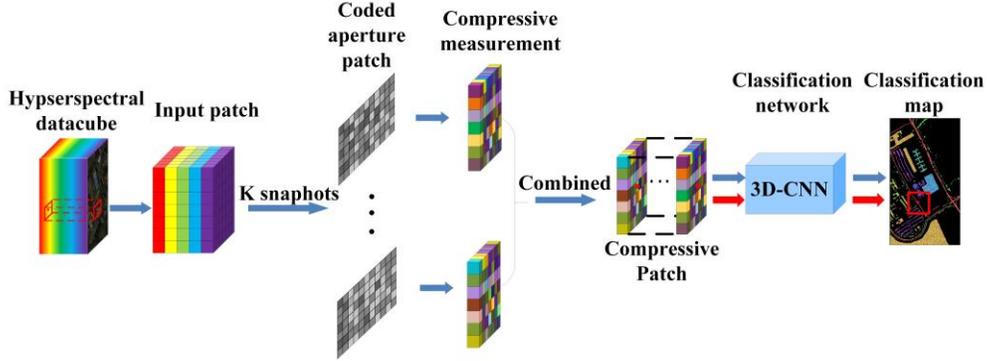

Fig. 5. Overview of the training process (blue arrows) and testing process (red arrows) of 3D-CCNN.

## 4. Experimental results

In this section, we evaluate the 3D-CCNN method on two public hyperspectral datasets, including the Pavia University dataset and the Salinas Valley dataset [7]. In addition, we compare the proposed method with several competitive methods, including the convolution neural network and support vector machine (SVM) classifier [32, 33]. In the comparative experiments, we use the random coded apertures or blue noise coded apertures in the DD-CASSI system. Note that the blue noise coding strategy has been proved to be optimal for the reconstruction in CASSI [34]. In this paper, the transmittance of all random coded apertures is set as 0.5. The blue noise coded apertures are generated based on the relevant method in [34]. In the practical implementation of the classification system, we need to calibrate and keep alignment between the coded aperture pattern and the detector. We can make a cross mark on the coded aperture, and use a standard whiteboard to replace the target. The cross mark can be imaged on the detector, and we can adjust the position of the cross-mark image to align the coded aperture with the detector. More details of the calibration method in CASSI system can be found in literature [35]. In addition, the modulation of coded aperture in real CASSI system cannot be regarded as ideal coding. Thus, we can first obtain the images of the coded apertures on the detector, and then use these images to calibrate the transmission functions of coded apertures.

The first four comparative methods are defined as follows:
**(1) "Rand-compress-3D-CNN" Method:** Use the random coded apertures to obtain the compressive measurements, and then perform the hyperspectral classification using the seven-layer 3D-CNN model described in Section 3.1.
**(2) "Bluenoise-compress-3D-CNN" Method:** Use blue noise coded apertures to obtain the compressive measurements, and then perform the hyperspectral classification using the seven-layer 3D-CNN model.

(3) **"Rand-compress-SVM" Method:** Use the random coded apertures to obtain the compressive measurements, and then perform the hyperspectral classification using the SVM classifier.
(4) **"Bluenoise-compress-SVM" Method:** Use the blue noise coded apertures to obtain the compressive measurements, and then perform the hyperspectral classification using the SVM classifier.

All of the methods mentioned above perform the classification in the compressive domain. It is noted that the hyperspectral data cube of the target scene can be reconstructed from compressive measurements by solving for an $l_1$-norm minimization problem. The details of the reconstruction methods have been published in literature [13, 36, 37]. It is natural to ask whether the classification accuracy can be improved by using the reconstructed hyperspectral data cube instead of the compressive measurements. To answer this question, we compare the proposed method with the following comparative methods:

(5) **"Rand-construct-3D-CNN" Method:** Use the random coded apertures to obtain the compressive measurements, and then use 3D-CNN to perform the classification based on the reconstructed hyperspectral data cube.
(6) **"Bluenoise-construct-3D-CNN" Method:** Use the blue noise coded apertures to obtain the compressive measurements, and then use 3D-CNN to perform the classification based on the reconstructed hyperspectral data cube.
(7) **"Rand-construct-SVM" Method:** Use the random coded apertures to obtain the compressive measurements, and then use SVM to perform the classification based on the reconstructed hyperspectral data cube.
(8) **"Bluenoise-construct-SVM" Method:** Use the blue noise coded apertures to obtain the compressive measurements, and then use SVM to perform the classification based on the reconstructed hyperspectral data cube.

Furthermore, in this paper the proposed method is also compared with the classifiers, where the original hyperspectral data cube is assumed to be available:

(9) **"Original-3D-CNN" Method:** Use 3D-CNN to perform the classification directly based on the original hyperspectral data cube of target scene.
(10) **"Original-SVM" Method:** Use SVM to perform the classification directly based on the original hyperspectral data cube of target scene.

In the following simulations, several indices are used to quantitatively assess the classification performance, including the overall accuracy (OA), average accuracy (AA), and Kappa coefficient ($K_a$). The OA is defined as the ratio of the correctly classified samples over all testing samples. The AA is the mean value of accuracy for each category. The $K_a$ is a statistical metric that provides mutual information regarding the agreement between the ground truth map and the classification map [7].

*4.1 Simulation result on Pavia University dataset*

The Pavia University dataset was collected by the Reflective Optics Imaging Spectrometer (ROSIS) over University of Pavia, Italy [7]. The spectral image in this dataset is characterized by high spatial resolution (1.3m per pixel) comprising $640 \times 340$ spatial pixels, and 103 spectral reflectance bands in the wavelength range from 0.43 $\mu m$ to 0.86 $\mu m$. In the following, a $256 \times 256 \times 103$ cube is truncated from the entire dataset to be used as the original hyperspectral data cube. Figure 6(a) shows the false-color composite image of the Pavia University spectral data. Figure 6(b) shows the ground truth of the classification map, which consists of nine distinct classes with different colors. Each class label corresponds to a different kind of objects in the urban cover, and the black regions represent the unlabeled pixels. From the image, 30% of the labeled pixels are randomly chosen to be used as the training samples, and the remaining 70% pixels are used for testing. Figure 7(a) illustrates one of randomly initialized coded aperture patterns, and Fig. 7(b) illustrates the optimized coded aperture pattern after joint optimization of 3D-CCNN. The difference pattern between the initial coded aperture and the optimized coded aperture is showed in Fig. 7(c).

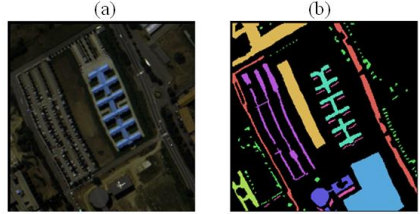

Fig. 6. (a) False-color composite image of the Pavia University spectral data and (b) ground truth of the classification map including nine distinct classes, where black regions represent the unlabeled pixels.

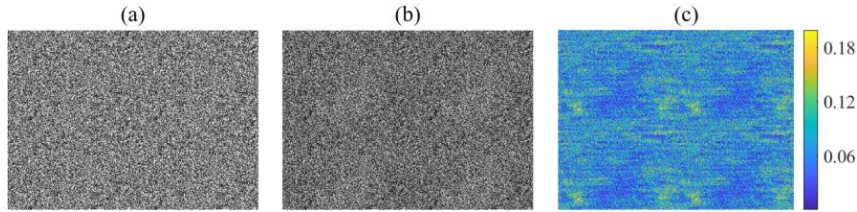

Fig. 7. Illustration of coded apertures for the Pavia University spectral data: (a) the initial coded aperture pattern, (b) the optimized coded aperture pattern using 3D-CCNN, and (c) the difference between the initial and optimized coded aperture patterns.

Figure 8 shows the classification results on the Pavia University dataset using the (a) proposed 3D-CCNN method, and the first four comparative methods, including the (b) Rand-compress-3D-CNN method, (c) Bluenoise-compress-3D-CNN method, (d) Rand-compress-SVM method, and (e) Bluenoise-compress-SVM method. The number of snapshots is 5, which means that the compression ratio of DD-CASSI is about 5%. In the 3D-CCNN framework, the spatial dimension of each patch in both training and testing sets is $7\times 7$. Thus, the patch size of the 3D-CNN input is $7\times 7\times 5$.

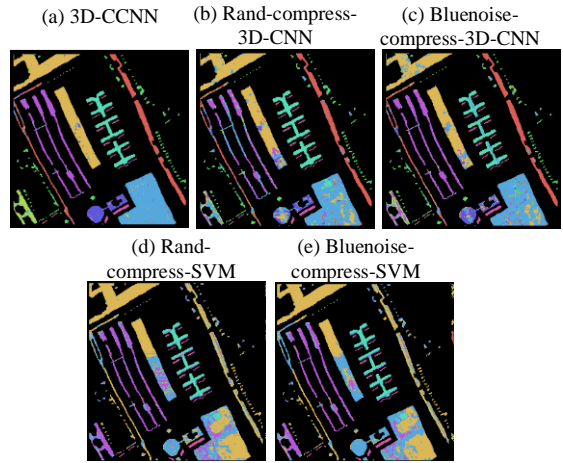

Fig. 8. Classification results on the Pavia University dataset using the following methods: (a) the proposed 3D-CCNN method, (b) Rand-compress-3D-CNN method, (c) Bluenoise-compress-3D-CNN method, (d) Rand-compress-SVM method, and (e) Bluenoise-compress-SVM method.

Table I shows the classification performance of the proposed method and the first four comparative methods using the Pavia University dataset. These metrics are calculated by averaging over several runs of the experiments. From the second row to the tenth row, it shows the percentage of accurate classification for each king of objects. The last three rows provide the OA, AA, and $K_a$ of the overall classification result.

**Table I. The classification performance of the proposed method and the first four comparative methods using the Pavia University dataset**

| Class | 3D-CCNN | Rand +compress +3D-CNN | Blue noise +compress +3D-CNN | Rand +compress +SVM | Blue noise +compress +SVM |
|---|---|---|---|---|---|
| Asphalt | 85.5% | 84.4% | 85.2% | 0.0% | 0.0% |
| Meadows | 94.6% | 80.5% | 80.1% | 53.4% | 54.4% |
| Gravel | 76.5% | 32.1% | 44.4% | 0.0% | 0.0% |
| Trees | 80.9% | 78.2% | 77.9% | 0.0% | 0.7% |
| Metal sheets | 94.5% | 94.8% | 88.6% | 81.1% | 82.4% |
| Bare soil | 87.6% | 69.0% | 72.7% | 34.5% | 35.9% |
| Bitumen | 87.6% | 52.4% | 60.7% | 0.0% | 0.0% |
| Bricks | 86.7% | 67.0% | 85.2% | 61.4% | 69.4% |
| Shadows | 93.3% | 98.7% | 96.3% | 99.8% | 99.8% |
| OA (%) | 86.58% | 73.39% | 76.34% | 45.78% | 47.55% |
| AA (%) | 87.47% | 73.01% | 76.79% | 36.69% | 38.07% |
| $K_a$ | 0.839 | 0.678 | 0.715 | 0.315 | 0.334 |

Above simulations show that the proposed 3D-CCNN method outperforms other methods directly based on the compressive measurements. The gain of the proposed method is mainly attributed to the joint optimization between coded apertures and network parameters. In addition, both of the 3D-CNN and SVM classifiers perform better on the blue noise coded apertures than the random coded apertures. That is because the blue noise coding strategy achieves more uniform sampling than the random ones, and is beneficial to capture more structure information from the target scene. Based on the same type of coded apertures, the 3D-CNN outperforms the SVM classifier, which proves the superior prediction capacity of the deep learning approach.

Figure 9 shows the classification results on the Pavia University dataset using the (a) Original-3D-CNN method, (b) Rand-construct-3D-CNN method, (c) Bluenoise-construct-3D-CNN method, (d) Original-SVM method, (e) Rand-construct-SVM method, and (f) Bluenoise-construct-SVM method. From the image, 30% of the labeled pixels are randomly chosen to be used as the training samples, and the remaining 70% pixels are used for testing. These methods perform the classification based on the original hyperspectral data cube or the reconstructed data cube. Table II provides the metrics of classification performance for these methods.

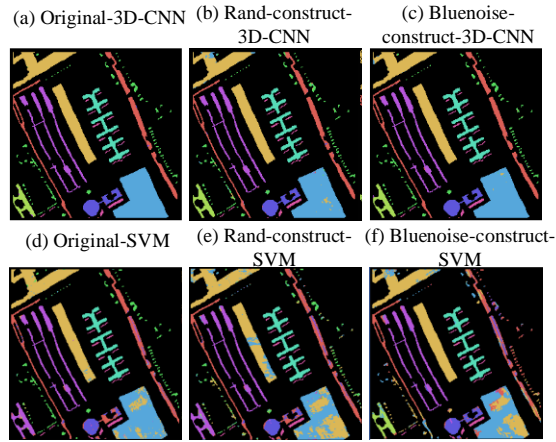

Fig. 9. Classification results on the Pavia University dataset using the following methods: (a) the Original-3D-CNN method, (b) Rand-construct-3D-CNN method, (c) Bluenoise-construct-3D-CNN method, (d) Original-SVM method, (e) Rand-construct-SVM method, and (f) Bluenoise-construct-SVM method.

From Fig. 9 and Table II, it is observed that the classifiers with blue noised coded apertures outperform those with random coded apertures, since the blue noise coding strategy achieves higher reconstruction quality than the random coding. Although the 3D-CNN applied on the reconstructed data cube outperforms the proposed 3D-CCNN method, it is important to remark the computational complexity to solve the reconstruction problem. In our simulations, the reconstruction process will take about 1168s. On the other hand, the 3D-CCNN only takes 57s to calculate the entire classification map. That is the proposed 3D-CCNN method will achieve more than 20-fold acceleration compared to the methods based on the reconstruction. What is more interesting, the performance of 3D-CCNN with only 5% compressive ratio is even better than the well-known SVM classifier that performs on the reconstructed full data cube.

**Table II. The classification performance of the last six comparative methods using the Pavia University dataset**

| Class | Original +3D-CNN | Rand +construct +3D-CNN | Blue noise +construct +3D-CNN | Original +SVM | Rand +construct +SVM | Blue noise + construct +SVM |
|---|---|---|---|---|---|---|
| Asphalt | 96.7% | 96.0% | 94.3% | 88.8% | 87.9% | 70.5% |
| Meadows | 96.7% | 93.9% | 94.9% | 91.0% | 73.7% | 79.4% |
| Gravel | 95.6% | 91.6% | 96.1% | 2.9% | 7.0% | 91.1% |
| Trees | 97.0% | 93.3% | 89.0% | 96.5% | 70.6% | 0.7% |
| Metal sheets | 100.0% | 99.9% | 99.8% | 99.7% | 99.3% | 99.0% |
| Bare soil | 96.3% | 93.1% | 93.9% | 90.3% | 65.8% | 72.3% |
| Bitumen | 97.6% | 96.0% | 96.0% | 71.9% | 72.1% | 80.4% |
| Bricks | 99.6% | 98.6% | 98.9% | 83.0% | 82.4% | 93.6% |
| Shadows | 97.6% | 97.4% | 97.9% | 100.0% | 99.6% | 99.8% |
| OA | 94.88% | 92.79% | 93.02% | 88.20% | 76.26% | 78.67% |
| AA | 97.46% | 95.53% | 95.64% | 80.46% | 73.16% | 76.31% |
| $Ka$ | 0.939 | 0.914 | 0.916 | 0.857 | 0.710 | 0.739 |

*4.2 Simulation result on Salinas Valley dataset*

Next, we test and evaluate all the classification methods on another dataset called Salinas Valley dataset. This dataset was acquired by the Airborne Visible/Infrared Imaging

Spectrometer (AVIRIS) on the Valley of Salinas, USA [7]. This spectral image exhibits a spatial resolution of 3.7*m* per pixel with 512×217 spatial pixels, and 192 spectral bands in the wavelength range from 0.24μm to 2.40μm. A 216×216×192 data cube is truncated from the entire dataset to be used in the experiments. Figure 10(a) shows the false-color composite image of the Salinas Valley dataset, and Fig. 10(b) shows the ground truth of the classification map including 8 distinct categories, each of which corresponds to a different type of crops. From the image, 30% of the labeled pixels are randomly chosen to be used as the training samples, and the remaining 70% pixels are used for testing.

Figure 11 shows the classification results on the Salinas Valley dataset using the (a) proposed 3D-CCNN method and the first four comparative methods, including the (b) Rand-compress-3D-CNN method, (c) Bluenoise-compress-3D-CNN method, (d) Rand-compress-SVM method, and (e) Bluenoise-compress-SVM method. The number of snapshots is 10, and the compression ratio of DD-CASSI is about 5%. The patch size of the 3D-CNN input is $7\times7\times10$. Table III provides the classification performance metrics for different methods in Fig. 11. These metrics are calculated by averaging over several runs of the experiments. It is noted that, the proposed 3D-CCNN outperforms other classification methods directly based on compressive measurements, and we can obtain similar conclusions as those obtained from Fig. 8 and Table I.

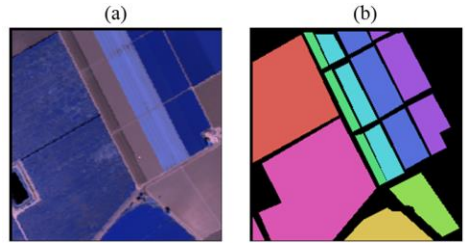

Fig. 10. (a) False-color composite image of the Salinas Valley spectral data, and (b) the ground truth of the classification map including nine distinct classes, where black regions represent the unlabeled pixels.

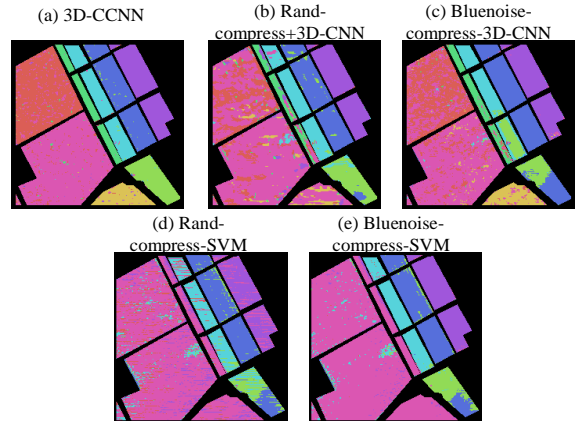

Fig. 11. Classification results on the Salinas Valley dataset using the following methods: (a) the proposed 3D-CCNN method, (b) Rand-compress-3D-CNN method, (c) Bluenoise-compress-3D-CNN method, (d)Rand-compress-SVM method, and (e) Bluenoise-compress-SVM method.

**Table III. The classification performance of the proposed method and the first four comparative methods using the Salinas Valley dataset**

| Class | 3D-CCNN | Rand +compress +3D-CNN | Blue noise +compress +3D-CNN | Rand +compress +SVM | Blue noise +compress +SVM |
|---|---|---|---|---|---|
| Broccoli green weeds | 87.4% | 49.4% | 65.2% | 19.5% | 1.0% |
| Fallow | 86.1% | 27.6% | 73.4% | 0.0% | 0.0% |
| Fallow rough plow | 97.9% | 85.1% | 51.8% | 53.4% | 73.7% |
| Fallow smooth | 97.1% | 72.1% | 89.9% | 0.0% | 0.0% |
| Stubble | 97.9% | 91.7% | 80.9% | 70.0% | 85.8% |
| Celery | 98.0% | 94.2% | 87.1% | 87.5% | 89.4% |
| Grapes untrained | 96.6% | 93.1% | 97.9% | 72.3% | 93.1% |
| Vineyard untrained | 90.1% | 69.0% | 75.9% | 59.9% | 63.6% |
| OA | 90.76% | 71.64% | 75.76% | 57.04% | 62.74% |
| AA | 93.89% | 72.78% | 77.76% | 45.33% | 50.83% |
| *Ka* | 0.887 | 0.649 | 0.703 | 0.451 | 0.521 |

Figure 12 shows the classification results on the Salinas Valley dataset using the (a) Original-3D-CNN method, (b) Rand-construct-3D-CNN method, (c) Bluenoise-construct-3D-CNN method, (d) Original-SVM method, (e) Rand-construct-SVM method, and (f) Bluenoise-construct-SVM method. The metrics of classification performance for these methods are provided in Table IV. From Fig. 12 and Table IV, we can obtain similar conclusions as those obtained from Fig. 9 and Table II. It is noted that the proposed 3D-CCNN with about 5% compression ratio even outperforms the well-known SVM classifier based on the full data cube. The research work and the revision are contributed by all of the four-authors.

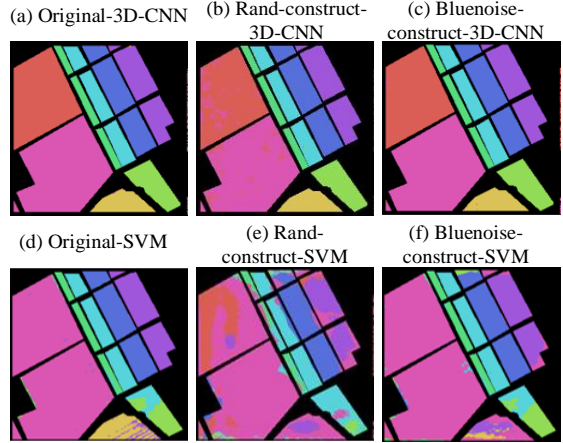

Fig. 12. Classification results on the Salinas Valley dataset using the following methods: (a) the Original-3D-CNN method, (b) Rand-construct-3D-CNN method, (c) Bluenoise-construct-3D-CNN method, (d) Original-SVM method, (e) Rand-construct-SVM method, and (f) Bluenoise-construct-SVM method.

**Table IV. The classification performance of the last six comparative methods using the Salinas Valley dataset**

| Class | Original+ 3D-CNN | Rand +construct +3D-CNN | Blue noise +construct +3D-CNN | Original +SVM | Rand +construct +SVM | Blue noise + construct +SVM |
|---|---|---|---|---|---|---|
| Broccoli green weeds | 97.5% | 91.7% | 97.5% | 0.0% | 53.6% | 7.9% |
| Fallow | 89.0% | 89.2% | 88.8% | 90.4% | 0.0% | 26.0% |
| Fallow rough plow | 99.9% | 99.5% | 99.9% | 76.5% | 75.2% | 71.0% |
| Fallow smooth | 99.6% | 99.6% | 99.3% | 96.1% | 0.0% | 94.2% |
| Stubble | 100.0% | 99.4% | 99.8% | 86.5% | 74.9% | 84.0% |
| Celery | 99.2% | 99.2% | 99.2% | 99.2% | 96.5% | 97.6% |
| Grapes untrained | 99.2% | 99.2% | 99.3% | 95.1% | 94.2% | 86.3% |
| Vineyard untrained | 98.0% | 93.8% | 97.9% | 72.9% | 62.5% | 69.0% |
| OA | 96.37% | 93.81% | 96.27% | 73.81% | 66.83% | 68.27% |
| AA | 97.80% | 96.45% | 97.71% | 77.09% | 57.11% | 67.00% |
| *Ka* | 0.956 | 0.921 | 0.955 | 0.671 | 0.580 | 0.598 |

## 5. Conclusion

This paper develops an efficient 3D-CCNN method to perform hyperspectral classification directly based on the DD-CASSI compressive measurements. The proposed 3D-CCNN method successfully avoids the time-consuming reconstruction procedure, and the influence of reconstruction artifacts. In addition, the hardware-based coded apertures and the software-based 3D-CNN are combined into a uniform framework, which are then jointly optimized by an end-to-end training method to increase the degrees of optimization freedom. Based on a set of simulations, the proposed 3D-CCNN is proved to outperform the 3D-CNN and SVM classifiers based on the compressive measurements. Also, the performance of 3D-CCNN with only about 5% compression ratio is comparable or even better than the SVM classifier based on the full data cube. This is the updated version of the previous articles. The contributions in the paper and the modifications in the revision have been provided by the corresponding author Xu Ma, and all of the other coauthors.

**Funding.**

Fundamental Research Funds for the Central Universities (2018CX01025, 2020CX02002).

**Disclosures.**

The authors declare no conflicts of interest.